# Design and Development of a Heart Rate Measuring Device using Fingertip

M.M.A. Hashem[1], Rushdi Shams[2], Md. Abdul Kader[3], and Md. Abu Sayed[4]
Department of Computer Science and Engineering
Khulna University of Engineering & Technology (KUET)
Khulna 9203, Bangladesh
{mma_hashem[1], rushdecoder[2], kdr2k4[3], sayed_cse_kuet[4]}@yahoo.com

*Abstract*— In this paper, we presented the design and development of a new integrated device for measuring heart rate using fingertip to improve estimating the heart rate. As heart related diseases are increasing day by day, the need for an accurate and affordable heart rate measuring device or heart monitor is essential to ensure quality of health. However, most heart rate measuring tools and environments are expensive and do not follow ergonomics. Our proposed Heart Rate Measuring (HRM) device is economical and user friendly and uses optical technology to detect the flow of blood through index finger. Three phases are used to detect pulses on the fingertip that include pulse detection, signal extraction, and pulse amplification. Qualitative and quantitative performance evaluation of the device on real signals shows accuracy in heart rate estimation, even under intense of physical activity. We compared the performance of HRM device with Electrocardiogram reports and manual pulse measurement of heartbeat of 90 human subjects of different ages. The results showed that the error rate of the device is negligible.

## I. INTRODUCTION

Heart rate indicates the soundness of our heart and helps assessing the condition of cardiovascular system [1]. In clinical environment, heart rate is measured under controlled conditions like blood measurement, heart voice measurement, and Electrocardiogram (ECG) [4] but it can be measured in home environment also [2]. Our heart pounds to pump oxygen-rich blood to our muscles and to carry cell waste products away from our muscles. The more we use our muscles, the harder our heart works to perform these tasks- means our heart must beat faster to deliver more blood. A heart rate monitor is simply a device that takes a sample of heartbeats and computes the Beats per Minute (bpm) so that the information can easily be used to track heart condition. There are two types of methods to develop heart monitors- electrical and optical methods. The electrical method has an average error of 1 percent and average cost of $150.00. The optical method has an accuracy rating of 15 percent and an average cost of $20.

The average resting human heart rate is about 70 bpm for adult males and 75 bpm for adult females. Heart rate varies significantly between individuals based on fitness, age and genetics. Endure athletes often have very low resting heart rates. Heart rate can be measured by measuring one's pulse. Pulse measurement can be achieved by using specialized medical devices, or by merely pressing one's fingers against an artery (typically on the wrist or the neck). It is generally accepted that listening to heartbeats using a stethoscope, a process known as auscultation, is a more accurate method to measure the heart rate [3]. There are many other methods to measure heart rates like Phonocardiogram (PCG), ECG, blood pressure wave form [5] and pulse meters [6] but these methods are clinical and expensive. There are other cost-effective methods that are implemented with sensors as proposed in [7] and [8] but they are susceptible to noise and movement of subject and artery.

In this paper, the design and development of a low powered HRM device is presented that provides an accurate reading of the heart rate using optical technology. The device is ergonomic, portable, durable, and cost effective. We incorporated the optical technology using standard Light Emitting Diode (LED) and photo-sensor to measure the heart rate within seconds using index finger. A microcontroller is programmed to count the pulse. The heart rate is digitally displayed on an LCD controlled by the same microcontroller that counts the pulse.

The organization of the paper is as follows. In Section II, we discuss the system overview. Section III describes the methodologies by which HRM device works. Section IV shows the experimental results. Section V concludes the paper.

## II. SYSTEM OVERVIEW

In this section, we discuss the system overview like pulse detection, signal extraction, pulse amplification, and physical properties of our propose HRM device.

### A. Pulse Detection

A pulse (heartbeat) detector in heart rate monitors consist of the two parts: a pulse sensing unit and a heart rate displaying unit [4]. Our device uses two red LEDs and a photo-sensor to measure ones heart rate through the change of blood reflectivity on the index finger. The power transmitted by the LEDs is matched with the photo sensor in such a way that the resistance will vary within the range of the photo sensor after attenuations through the index finger. Since attenuations vary depending on the person using the device, our specifications assume that the attenuation is, 80 percent, on average, of the light transmitted. A resistance network is used with the sensor to transform the changes in resistance to the changes in voltage. The voltage created varies between 0 and 10 mV with respect to each heart pulse. Fig. 1 shows a clip sensor which consists of two high intensity LEDs that illuminate the tissue and a Light Detective Resistor (LDR) whose resistance changes according to the amount of light transmitted from the tissue.

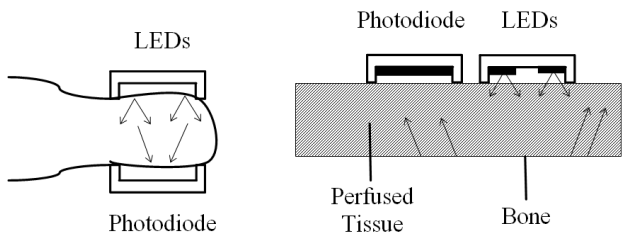

Figure 1.  Finger positioning on the HRM device

The LED and LDR are mounted in a spring loaded device that can be clipped into the fingertip. The light emitted by the LED is diffusely scattered through the fingertip tissue. An LDR or photo sensor positioned on the surface of the skin on the opposite side can measure light transmitted through it. Light is absorbed well in blood and weakly absorbed in tissue. Any changes in blood volume will be registered since increasing (or decreasing) volume will cause more or less absorption. Assuming the subject does not move the level of absorption of the tissue and non-pulsating fluids remains same.

*B. Signal Extraction*

We used a band pass filter to remove any interference caused by ambient light and level detection distortions. The filter used will have a cutoff frequency of 2.5 Hz to allow a maximum heart rate of 125 bpm to be measured by the device with accuracy. This roll off provides an attenuation of 60Hz, by 23.5dB. The pulse has a –14 dB Signal to Noise Ratio (SNR) before it passes through the filter. The pass-band frequencies are amplified by a factor of 40 dB with a small signal amplifier. We used DC blocking to prevent immeasurable pulses caused by a high DC offset from ambient light.

*C. Pulse Amplification*

The extracted signal is analyzed by an amplifier to provide a pulse of high amplitude to be fed into the microcontroller input. The amplifier detects the peak of each pulse and creates a corresponding pulse of high amplitude. This stage of the design requires that the amplified and filtered heart pulse signal have a SNR of 20 dB to obtain a clean pulse of high amplitude. The time between each successive rising of high amplitude pulse edge is interpreted by the microcontroller as the period between each heart pulse.

To detect signal amplification an LM358 is used and the heart signal is amplified twice after passing through band pass filter (Fig. 2). Finally, we get the amplified output that the microcontroller uses it as its input and calculates the heart rate.

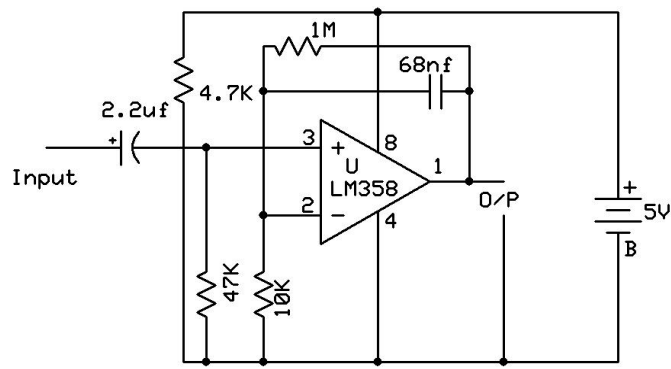

Figure 2.  Portion of the pulse amplifier circuit [9]

*D. Physical Properties*

The heart rate is displayed on an LCD controlled by a microcontroller. As improper finger placement may take place, a flashing segment on the LCD will be used to help the user identify when a heart rate is measured. A character LCD is used to show the formatted result of heart rate. The accuracy of our device is controlled primarily by the amount of time used to average the pulse rate from the user. The more time used means more samples and thus greater accuracy. Our device uses a method that has a dynamic time for pulse gathering. The pulse is averaged after a set number of pulses are obtained, five in our case, by the microcontroller. This allows a better distribution of accuracy over the range of pulses that can be measured by the device.

The device operates using a 5 Volt battery source because of the small packaging. The battery should last one year under normal use; however, this will vary based on the use. Typical usage of 15 minutes per day will allow the device to operate for two months. The device is designed to operate in a standard temperature environment of $-30^0$ C to $80^0$ C. The package is small and lightweight and hence portable. The final package dimensions is no larger than 3.5" x 2" x 1" (H x W x D). Cost of HRM device is kept to a minimum to maintain a competitive edge with currently available products. A maximum estimate for components at this point is $20. This cost is achieved through the unique design approach and component selection we employed. The physical properties of the HRM device are shown in Fig. 3.

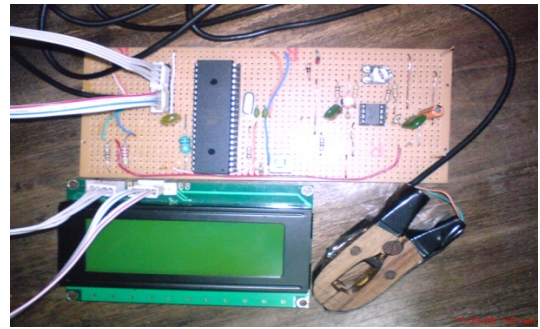

Figure 3.  Physical appearance of HRM device

## III. METHODOLOGY

### A. Optical Transmitter and Receiver Circuit

HRM measures the pulse rate through changes of blood flow through an index finger. Each pulse of blood from the heart increases the density of blood in the finger pulsatile tissue and causes a decrease in light power received by the photo-sensor. The photo-sensor does not pick up a purely AC signal as there are some DC components received from other non-pulsatile tissues and ambient light levels. The varying light levels received are converted into a varying resistance in the photo-sensor. The varying resistance is converted into a varying voltage by using a resistance network and power source. In order to do this, two red LEDs are used in combination with a photo-sensor to detect and transmit the pulse rate. Since the tissue in the human body acts a filter for red light, two red LEDs were chosen to allow the maximum amount of light energy to pass through the index finger. The circuit shown in Fig. 4 is designed to achieve this.

The circuit in Fig. 4 shows our pulse-rate to voltage converter and our source of constant red light that is designed and constructed to gather real-world data. The red LED is forward biased through a resistor to create a current flow. The value of R2 is chosen in a way so that it produces a maximum amount of light output. The calculated value is approximated to a resistance value that is commonly available. The photo-resistor is placed in series with a resistor to reduce the current drawn by the detection system and to prevent shorting the power supply when no light is detected by the photo resistor.

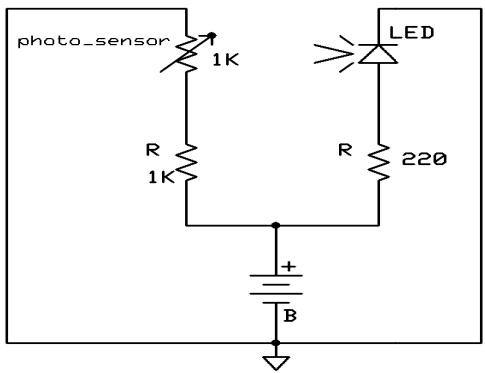

Figure 4. Pulse receiving circuit used in HRM device [9]

### B. Amplification of Pulse Rate Signal

To let the microcontroller counting the pulse rate, the signal must be amplified. An amplifier is used to find rising edges of the filtered signal received by the photo-sensor. This allows one pin of the microcontroller to be used as an input. The time between rising pulse edges is determined by the microcontroller so that the frequency of the heart rate can be measured. The designed circuit is shown in Fig. 5.

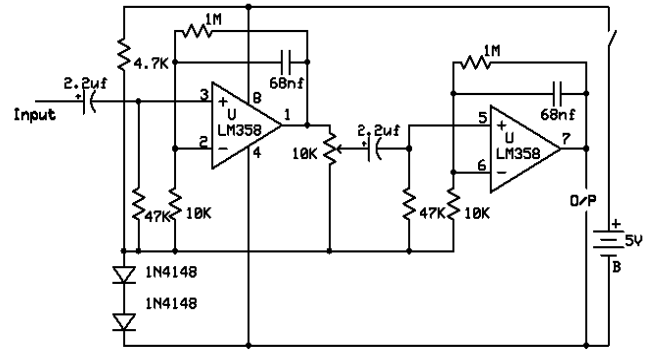

Figure 5. Amplifier Circuit used in HRM device [9]

The amplifier uses an LM358 dual op amp to provide two identical broadly-tuned band pass stages with gains of 100. Again, the type of op amp is not particularly critical, as long as it will work at 5V and drive the output rail to rail. The signal frequencies are boxed in by movement artifacts at the low end (generated by the peg moving and distorting the underlying tissues; light pegs are better) and at the top end by mains-hum interference. The circuit runs from a single 5 Volt battery and the output zero is offset by about 1 Volt by referring everything to an internal common line at a voltage set by a pair of forward-biased silicon diodes. This is convenient for interfaces with a 0-5 Volt input. The potentiometer allows the overall gain to be adjusted so as to prevent clipping on large signals. Components are not critical but the two 2.2 μF capacitors must be able to stand some reverse bias so they should be non-polarized or tantalum. The circuit can easily be made up on a small piece of strip board.

### C. Microcontroller

A microcontroller is an economical means of counting the pulse rate and controlling a LED display. The method used below allows the displays to be driven without the use of a display driver. The displays are set and refreshed by multiplexing the segment lines to the same I/O pins on the microcontroller.

Programming the microcontroller involves developing a calculation algorithm to count the pulse rate. The calculation algorithm for counting the pulse rate will be easy to develop using Firmware. The microcontroller will continuously be checking if a signal is fed into it. Once a signal is detected, the algorithm will begin according to the flow chart in Fig. 6.

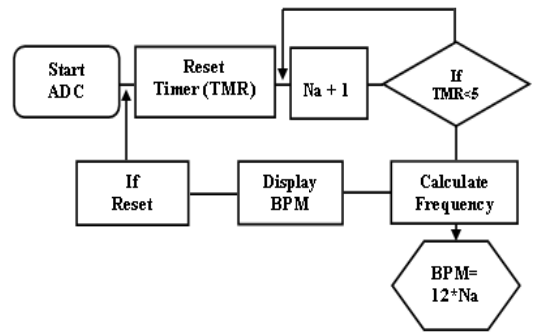

Figure 6. Algorithm for detecting pulse rate

Once each stage of the design has been simulated to prove their unit efficiency, they are integrated and in Fig. 7, the working principle is illustrated.

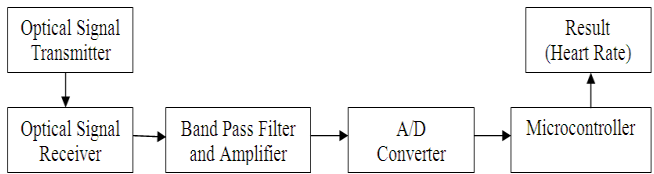

Figure 7. Working principle of HRM device

## IV. EXPERIMENTAL RESULTS

The first phase of the device, the optical receiver and transmitter, is constructed and tested. The output of the receiver is connected to an O-scope to obtain the heartbeat signal. Fig. 8 shows the heartbeat signal obtained by the device for a person using two different channels.

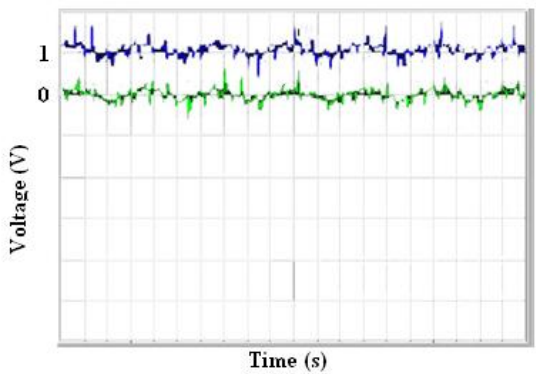

Figure 8. Extracted heartbeat signal

A band pass filter is used to filter the noise from the heartbeat signal. Fig. 9 shows the output obtained after removing the noise in the heartbeat signal.

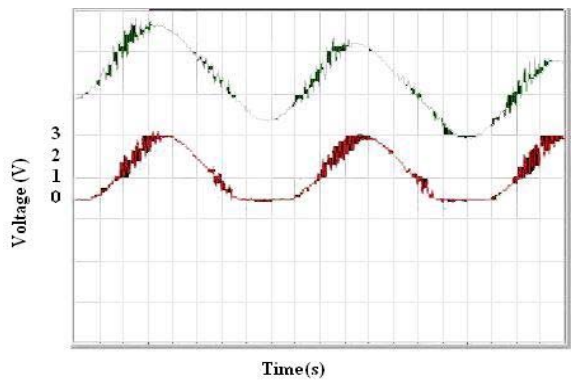

Figure 9. Filtered output signal

Comparing Fig. 9 and Fig. 10, we see that the high frequency noise of 120 Hz from ambient lights has been filtered out as expected. The filtered signal is required to have a SNR of 20 dB or greater, to ensure that the amplifier is able to correctly convert the continuous signal to a higher amplitude signal form without producing false trigger due to noise. The filtered signal has a SNR of approximately 24 dB, and this allows the amplifier to properly amplify the heartbeat. This test shows that the filter is able to remove high frequency noise from the heartbeat signal.

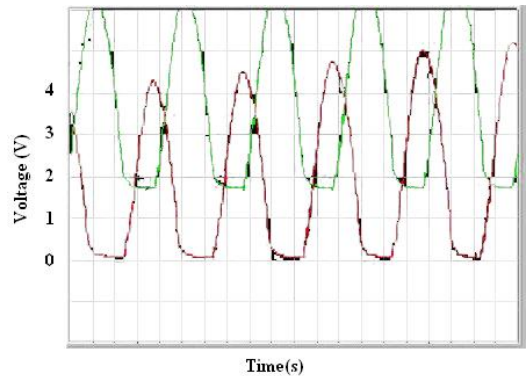

Figure 10. Amplified output signal

The microcontroller is programmed to count the number of peaks of the input signal in 5 or 10 seconds, and the result is further multiplied correspondingly by 12 or 6 to obtain the total number of peaks per minute. The LCD is connected to the microcontroller and a known frequency pulse signal is fed into it. The correct number of peaks per minute value is showed on the LCD. When the microcontroller is integrated into the entire design circuitry, it is able to count the number heartbeats per minute and drive the LCD to display the counted value.

The performance of HRM device is tested with the output of Electrocardiogram (ECG) for 10 patients. The error rate is calculated using (1)-

$$E = [100 \times |A - M|] \div A \tag{1}$$

Here, $A$ = Actual heart rate

$M$ = Measured heart rate and

$E$ = Error rate

The comparison shows that the HRM device has accuracy with a mean of 4.31 and standard deviation of 2.87. The comparison is shown in Table I.

TABLE I. ACCURACY COMPARISON WITH AN ECG.

| Electrocardiogram (bpm) | HRM device (bpm) | Error Rate (%) |
|---|---|---|
| 76 | 78 | 2.56 |
| 78 | 78 | 0 |
| 76 | 72 | 5.56 |
| 82 | 84 | 2.38 |
| 83 | 84 | 1.19 |
| 85 | 90 | 5.56 |
| 77 | 84 | 8.33 |
| 79 | 84 | 5.95 |
| 89 | 96 | 7.29 |
| 88 | 90 | 2.22 |

However, this accuracy may defer depending on the circumference size of the finger of the user. We also measured the error rate depending on the finger size and found that HRM works well with medium-sized fingers. The result is shown in Table II.

TABLE II. ACCURACY COMPARISON WITH DIFFERENT FINGER SIZE.

| Finger Circumference size | Error Rate (%) |
|---|---|
| Big Finger (3.0") | 15.67 |
| Medium Finger (2.5") | 4.31 |
| Small Finger (2.125") | 8.91 |

The accuracy of the heart monitor is also tested manually as we selected 90 human subjects from different ranges of age from 3 years to 65 years, and measured their heartbeat manually from the pulse and with the HRM device. From Fig. 10, we see that the difference between actual heart rate and measured heart rate is small (e.g., error rate is low).

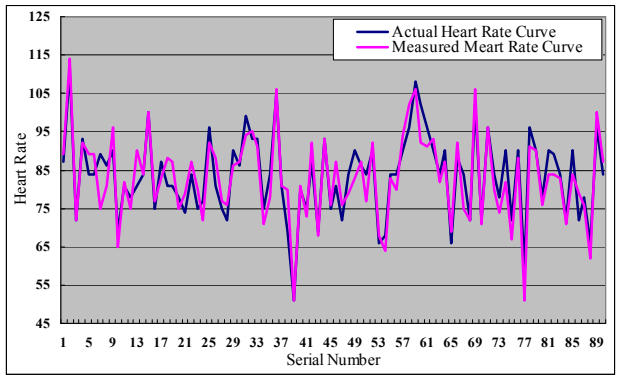

Figure 11. Difference between original heartbeat and measured heartbeat

Fig. 10 shows that the HRM device has average error rate of only 4.56 percent and hence depicts its efficiency in measuring heartbeats in a cost-effective and ergonomic manner.

## V. CONCLUSIONS

In this paper, the design and development of a Heart Rate Measuring device is presented that measures the heart rate efficiently in a short time and with less expense without using time consuming and expensive clinical pulse detection systems. Both analog and digital signal processing techniques are combined to keep the device simple and to efficiently suppress the disturbance in signals. Simulations showed that heart rate can be detected from changes of blood flow through an index finger. Experimental results showed that the heart rate can be filtered and digitized so that it can be counted to calculate an accurate pulse rate. The device is able to detect, filter, digitize, and display the heartbeat of a user ergonomically.

## REFERENCES

[1] R.G. Landaeta, O. Casas, and R.P. Areny, "Heart rate detection from plantar bioimpedance measurements", *28th IEEE EMBS Annual International Conference*, USA, 2006, pp. 5113-5116.

[2] P. F. Binkley, "Predicting the potential of wearable technology", *IEEE Eng. Med. Biol. Mag.*, Vol. 22, 2003, pp. 23-27.

[3] Wikipedia, "Heart rate", Available at: http://en.wikipedia.org/wiki/Heart_rate [December 27, 2009]

[4] H. Shim, J.H. Lee, S.O. Hwang, H.R. Yoon, and Y.R. Yoon, "Development of heart rate monitoring for mobile telemedicine using smartphone", *13th International Conference on Biomedical Engineering (ICBME 2008)*, Singapore, 2008, pp. 1116-1119.

[5] C. C. Tai and J.R.C. Chien, "An improved peak quantification algorithm for automatic heart rate measurements", *IEEE 27th Annual Conference on Engineering in Medicine and Biology*, China, 2005, pp. 6623-6626.

[6] Y. Chen, "Wireless heart rate monitor with infrared detecting module," US2005075577-A1, 2005.

[7] T. Usui, A. Matsubara, and S. Tanaka, "Unconstrained and noninvasive measurement of heartbeat and respiration using an acoustic sensor enclosed in an air pillow," *SICE 2004 Annual Conference*, 2004, vol. 3, pp 2648-2651.

[8] S. Rhee, B.-H. Yang, and H. H. Asada, "Modeling of finger photoplethysmography for wearable sensors," *21st Annual Conference and the 1999 Annual Fall Meeting of the Biomedical Engineering Soc. BMES/EMBS Conference*, 1999.

[9] Pico Technology, "Calculating the heart rate with a pulse plethysmograph", Available at: http://www.picotech.com/experiments/calculating_heart_rate/index.html [December 27, 2009]